\documentclass[pra, aps, twocolumn, groupedaddress, showpacs, superscriptaddress]{revtex4-2}
\usepackage{amssymb,amsmath,amsthm,color,graphicx,times,graphicx}\usepackage[caption=false]{subfig}
\usepackage{hyperref}
\usepackage{enumerate}
\usepackage{graphicx}
\usepackage{dsfont}
\usepackage{times}
\usepackage{bbold}
\usepackage{color}

\newcommand{\bra}[1]{\langle{#1}|}
\newcommand{\ket}[1]{|{#1}\rangle}

\usepackage{orcidlink}
\usepackage{hyperref}
\hypersetup{
	colorlinks=true,
	linkcolor=blue,
	filecolor=blue,      
	urlcolor=blue,
	citecolor=violet,
	pdfauthor={Yassine Dakir, Abdallah Slaoui, A-B A. Mohamed, Rachid Ahl Laamara and H. Eleuch},
	pdftitle={},
	pdfcreator={Abdallah Slaoui},
}

\providecommand{\openone}{\leavevmode\hbox{\small1\kern-4.3pt\normalsize1}}

\theoremstyle{plain}

\theoremstyle{definition}

\begin{document}
\title{Quantum teleportation and dynamics of quantum coherence and metrological non-classical correlations for open two-qubit systems: A study of Markovian and non-Markovian regimes}

\author{Yassine Dakir}\affiliation{LPHE-Modeling and Simulation, Faculty of Sciences, Mohammed V University in Rabat, Rabat, Morocco.}
\author{Abdallah Slaoui \orcidlink{0000-0002-5284-3240}}\email{Corresponding author: abdallah.slaoui@um5s.net.ma}\affiliation{LPHE-Modeling and Simulation, Faculty of Sciences, Mohammed V University in Rabat, Rabat, Morocco.}\affiliation{Centre of Physics and Mathematics, CPM, Faculty of Sciences, Mohammed V University in Rabat, Rabat, Morocco.}
\author{Abdel-Baset A. Mohamed}\affiliation{Department of Mathematics, College of Science and Humanities in Al-Aflaj, Prince Sattam bin Abdulaziz University, Saudi Arabia.}\affiliation{Department of Mathematics, Faculty of Science, Assiut University, Assiut, Egypt.}
\author{Rachid Ahl Laamara}\affiliation{LPHE-Modeling and Simulation, Faculty of Sciences, Mohammed V University in Rabat, Rabat, Morocco.}\affiliation{Centre of Physics and Mathematics, CPM, Faculty of Sciences, Mohammed V University in Rabat, Rabat, Morocco.}
\author{Hichem Eleuch}\affiliation{Department of Applied Physics and Astronomy, University of Sharjah, Sharjah 27272, United Arab Emirates.}\affiliation{College of Arts and Sciences, Abu Dhabi University, Abu Dhabi 59911, United Arab Emirates.}\affiliation{Institute for Quantum Science and Engineering, Texas A\&M University, College Station, TX 77843, USA}

\begin{abstract}
We investigate the dynamics of non-classical correlations and quantum coherence in open quantum systems by employing metrics like local quantum Fisher information, local quantum uncertainty, and quantum Jensen-Shannon divergence. Our focus here is on a system of two qubits in two distinct physical situations: the first one when the two qubits are coupled to a single-mode cavity, while the second consists of two qubits immersed in dephasing reservoirs. Our study places significant emphasis on how the evolution of these quantum criterion is influenced by the initial state's purity (whether pure or mixed) and the nature of the environment (whether Markovian or non-Markovian). We observe that a decrease in the initial state's purity corresponds to a reduction in both quantum correlations and quantum coherence, whereas higher purity enhances these quantumness. Furthermore, we establish a quantum teleportation strategy based on the two different physical scenarios. In this approach, the resulting state of the two qubits functions as a quantum channel integrated into a quantum teleportation protocol. We also analyze how the purity of the initial state and the Markovian or non-Markovian regimes impact the quantum teleportation process.

\par
\vspace{0.25cm}
\end{abstract}
\date{\today}

\maketitle
\section{Introduction}
One of the most interesting topics in the microscopic world involving quantum mechanics is quantum entanglement (QE), a particular property of quantum systems with no classical analogue \cite{Einstein1935,Bell1964,Hill1997}. The concept of QE shows how two or more objects can be bound to each other contrary to what our common sense would predict. If an ensemble of entangled particles is individually measured, then while the particles are physically separated by a large distance, in some ways they behave as a single object rather than as two separate objects and the resulting results may exhibit "non-local" effects \cite{Genovese2019}. Indeed, QE has been seen as a concrete physical resource with many important applications \cite{Peres1997,Fuchs}, ranging from cryptography \cite{Ekert1991} to teleportation \cite{Bennett1993,Asjad2021,Kirdi2023} to quantum computers \cite{Ladd2010,Boudreault2022,Nielsen}. Thus was born the quantum information sector which developed surprisingly. Besides, the further growth of the joint entanglement of a pair of qubits exposed to local noisy environments has recently attracted a lot of attention. The reason is related to the discovery of Yu and Eberly \cite{Yu2004} that, quite unexpectedly, the Markovian dynamics of joint entanglement of qubits and decoherence of a single qubit for this system can be quite different. In fact, quantum systems can be classified into two types; closed and open. A closed quantum system is completely isolated from its environment and does not interact with it, while an open quantum system exchanges energy, matter or information with its external environment. Over time, an open quantum system can undergo decoherence, which results in a loss of coherence between the system and its environment, and a loss of information about the initial state of the system.\par

The investigations concerning the decoherence started mainly through the works of Zeh \cite{Zeh2000} and Zurek \cite{Zurek2003}, with the initial aim of proposing a mechanism that could explain the emergence of classical properties from a purely quantum context. Furthermore, decoherence theory allows us to understand the evolution of correlations existing in a quantum system and could therefore help us to develop strategies to combat noise effects. Overcoming decoherence is a prerequisite for the very attractive idea of coding information in quantum systems. For this purpose, the possibility of QE disappearing completely at limited times has attracted most of the interest. This phenomenon of "entanglement sudden death" has been demonstrated in a quantum optics experiment \cite{Almeida2007}. Further, we keep discovering new situations in which dissipation play a crucial role in manipulating open quantum systems. In many circumstances, it is no longer possible to assume a significant boundary between the time scales of the system and the environment, resulting in non-markovian behavior and ultimately feedback from the environment to the system \cite{Breuer2016,Vega2017}. It is therefore essential to create a description of the system-environment interaction that is accurate and efficient. In this present work, we focus specifically on two scenarios of open quantum systems; The first one concerns a Jaynes-Cummings model (JCM) \cite{Jaynes1963} and the second scenario deals with a non-markovian dephasing model \cite{Daffer2004}. Basically, the JCM extensively employed in quantum optics, which serves as a fundamental framework for elucidating the interaction between two-level atoms and quantized electromagnetic fields in the dipole approximation. This model can be precisely solved when applying the rotating wave approximation, wherein non-energy-conserving terms are eliminated from the Hamiltonian. Consequently, the JCM has been a focal point of considerable theoretical and experimental investigations in recent decades. On the other hand, non-markovian dephasing model are designed to capture the effects of memory in the evolution of an  quantum system. Non-markovian dynamics can capture the implications of memory in the evolution of an open quantum systems and can be used to study a wide range of phenomena, including quantum memory and entanglement dynamics \cite{Breuer2016,Zidan2021,ElEuch2021,ElEuch2023,Vega2017}. The differentiation between markovian and non-markovian models lies in how they depict the temporal evolution of the system. A markovian model assumes that the quantum state of the system at a particular moment is solely determined by its immediate preceding state. Consequently, the properties of the system at time $t+1$ can be completely ascertained by knowing its state at time $t$. In other words, the Markovian model postulates that the future is independent of the past, given the system's current state. Conversely, a non-markovian model incorporates longer-term influences and permits non-local dependencies over time \cite{Wolf2008}. This implies that the quantum state of the system at a specific moment may depend on its past states over several time intervals, rather than solely relying on its immediately preceding state.\par

The characterization of non-classical correlations poses a significant challenge in this research field, and numerous measures have been introduced in the literature to quantify them in quantum systems \cite{Vedral2002,Modi2012,Horodecki2009}. Measures such as entanglement of formation and concurrence serve as quantitative indicators of entanglement \cite{Wootters1998}. Another measure, known as quantum discord (QD), based on the von Neumann entropy, provides a means to assess quantum correlations in bipartite quantum systems, even those that are separable. It was initially introduced by Ollivier and Zurek \cite{Ollivier2001} and Henderson and Vedral \cite{Henderson2001}. However, computing QD based on von Neumann entropy is generally a complex task. In fact, it has been proven to be NP-complete \cite{Huang2014}, and only partial results are available for certain special two-qubit states. In response to these obstacles, scientists have put forth a geometric approach for quantifying quantum discord that employs Schatten $p$-norms. The initial formulation of this geometric quantum discord measure was introduced in \cite{Dakic2010}, where the Hilbert-Schmidt norm was employed. Despite being computationally tractable \cite{Bellomo22012,Brown2012,Bellomo2012}, the geometric QD is not an ideal indicator of non-classical correlations. This occurs because the geometric QD, measured using the Hilbert-Schmidt distance, can actually increase when local quantum operations are performed on the unmeasured qubit \cite{Piani2012}. This undesirable characteristic stems from the absence of contractivity which is a vital property for any quantum correlations quantifier. It is now established that the trace norm is the only Schatten $p$-norm suitable for the geometric measure of QD \cite{Paula2013,ABMohamed2019,SlaouiM2018,Paula2013}. Recently, a new measure called local quantum uncertainty (LQU) \cite{Girolami2013} has been introduced to examine pairwise quantum correlations of the QD type in multipartite systems. It possesses all the desirable properties expected of a reliable quantum correlations quantifier. This measure is based on the concept of Skew information, initially introduced by Wigner to determine the uncertainty in measuring an observable \cite{Wigner1963,Luo2004}. The advantage of the LQU is its analytical computability for any qubit-qudit system. Furthermore, it is worth noting that this new quantum correlation quantifier is closely linked to quantum Fisher information, which is commonly employed in the field of quantum metrology \cite{SlaouiB2019}. In a recent study \cite{Kim2018}, the introduction of local quantum Fisher information (LQFI) provided a means to quantify non-classical correlations using QFI. This powerful quantifier, akin to quantum discord, involves minimizing QFI over a locally informative observable related to a specific subsystem. Moreover, LQFI holds great potential as a tool for comprehending the impact of quantum correlations, beyond entanglement, in improving the precision and efficiency in quantum estimation protocols.\par

The purpose of this study is to investigate the behavior of non-classical correlations (LQU and LQFI) and quantum coherence in two distinct open quantum systems. The first system consists of two qubits coupled with a single-mode cavity field, and it is analyzed under the rotating wave approximation. The second system involves two qubits interacting with dephasing reservoirs. A comparative analysis of their time evolution is conducted, and particular attention is given to examining the impact of both Markovian and non-Markovian environments on their quantumness. Both models share the same dependence on the initial state, enabling a meaningful comparison between them. Additionally, we investigate the quantum teleportation of a two-qubit system in an arbitrary pure entangled state through the two considered models as a quantum channel \cite{Yu2008,Liu2006}. The output average fidelity is examined to validate the effectiveness of the teleportation process. This article is organized as follows: Sec.\ref{Sec2} provides an overview of LQFI, LQU, and QC, which are used to quantify the degree of quantumness in our models of interest. Sec.\ref{Sec3} is dedicated to analyzing the behavior of LQU, LQFI, and QC in the Jaynes-Cummings model as well as in a non-Markovian dephasing model. Furthermore, in Sec.\ref{Sec4}, we discuss the success of quantum teleportation in both models by measuring the average fidelity. Finally, we sum up our results in Sec.\ref{Sec5}.

\section{Quantifiers of non-classical correlations}\label{Sec2}
\subsection{Non-classical correlations measured by local quantum uncertainty}
Using local measurements on a part of the given quantum state $\rho_{AB}$, local quantum uncertainty (LQU) is a quantifier of non-classical correlations that captures a fully quantum part of the bipartite state $\rho_{AB}$ \cite{Girolami2013} and which goes beyond entanglement. It can be computed as
\begin{equation}
	{\cal U}_{Q}\left(\rho_{AB}\right)= \min_{{\cal K}_{A}} {\cal I}\left(\rho_{AB},{\cal K}_{A}\otimes \openone_{B}\right), \label{LQU}
\end{equation}
with the minimum is optimized over all local observables ${\cal K}_{A}$ on the subsystem $A$, and ${\cal I}\left(\rho,{\cal K}_{A}\otimes \openone_{B}\right)$ is the Wigner–Yanase skew information \cite{Wigner1963,Luo2003} that quantifies the information content and the noncommutativity in the state $\rho_{AB}$ with regard to the local observable ${\cal K}_{A}$, it writes as
\begin{equation}
	{\cal I}\left(\rho_{AB},{\cal K}_{A}\otimes \openone_{B}\right)= -\frac{1}{2}{\rm Tr}\left( \left[\sqrt{\rho_{AB}},{\cal K}_{A}\otimes \openone_{B}\right]^{2}\right).
\end{equation} 
Therefore, the minimum quantum uncertainty associated with a single measurement on a single subsystem is referred to as LQU. Importantly, LQU is one of the most adapted quantifiers of non-classical correlations and has been shown to satisfy all the necessary physical requirements for such a discord-like quantifier. For instance, LQU is non-increasing under local operations on B, it vanishes if and only if the quantum state is a zero-discord state, it is invariant under local unitary operations for bipartite quantum systems, and lastly, it reduced to a monotone entanglement for pure states. For bipartite $2\otimes d$-quantum states, a closed from of LQU can be derived as
\begin{equation}
{\cal U}_{Q}\left(\rho_{AB}\right)=1-\Lambda_{\max}\left({\cal W}\right),\label{lqu}
\end{equation}
$\Lambda_{\max}$ is the largest eigenvalue of the $3\times3$ symmetric matrix ${\cal W}\equiv[{\cal W}_{ij}]$, whose components are calculated using the following formula
\begin{equation}
	{\cal W}_{ij}={\rm Tr}\left\lbrace \sqrt{\rho_{AB}}\left( \sigma_{i}\otimes \openone_{B}\right) \sqrt{\rho_{AB}}\left(\sigma_{j}\otimes \openone_{B}\right)\right\rbrace, \label{wij}
\end{equation}
with $\sigma_{i,j}$ (where $i,j=1,2,3$) are the standard Pauli matrices acting on the part $A$.
\subsection{Non-classical correlations quantified by local quantum Fisher information}
Here we employ local quantum Fisher information (LQFI) as a second quantifier of non-classical correlations based on quantum uncertainty. Briefly, QFI $\mathcal{F}$ is the most commonly applied quantity to characterize the ultimate accuracy in the parameter estimation protocols via the quantum Cramér-Rao bound, where $\Delta\theta\geq1/\sqrt{\mathcal{F}}$ \cite{Paris2009,Szczykulska2016}. Recently, numerous efforts have been done to establish the relevance of non-classical correlations in quantum metrology, resulting in showing how quantum discord can significantly improve the precision of estimated parameter $\theta$ in unitary evolution processes, i.e in the case of unitary dynamics $\rho_{\theta}={\cal U}_{\theta}^{\dagger}\rho{\cal U}_{\theta}$ with ${\cal U}_{\theta}=\exp[i{\cal K}\theta]$. According to estimation theory, we can derive the QFI by optimizing the classical FI over all possible measurements and can be reformulated as
\begin{equation}
	\mathcal{F}\left( \rho_{\theta}\right)=\frac{1}{4}{\rm Tr}\left(\rho_{\theta}{\cal L}_{\theta}^{2}\right), \label{eq4}
\end{equation}
where ${\cal L}_{\theta}$ is the symmetric logarithmic derivative (SLD) associated to the estimated parameter $\theta$, which was determined by solving the equation
\begin{equation}
	\frac{\partial\rho_{\theta}}{\partial\theta} = \frac{1}{2}\left[\rho_{\theta}{\cal L}_{\theta} + {\cal L}_{\theta}\rho_{\theta}\right]. \label{ll}
\end{equation}
Applying this result to the spectrum decomposition $\rho_{\theta}=\sum_{i=1}^{N}\eta_{i}\ket{\psi_{i}}\bra{\psi_{i}}$ with $\eta_{i}\geq0$ and $\sum_{i=1}^{N} \eta_{i}=1$, the SLD can be computed explicitly as
\begin{equation}
{\cal L}_{\theta}=2\sum_{i,j}\frac{\left\langle\psi_{j}\right|\partial_{\theta}\rho_{\theta}\left| \psi_{i}\right\rangle}{\eta_{i}+\eta_{j}}\left|\psi_{j}\right\rangle\left\langle \psi_{i}\right|, \label{eq6}
\end{equation}
and the explicit formula for the QFI is obtained by reporting Eq. (\ref{eq6}) into Eq.(\ref{eq4}) as 
\begin{equation}
\mathcal{F}\left(\rho_{\theta}\right)=\frac{1}{2}\sum_{i \ne j} \frac{\left(\eta_{i}-\eta_{j}\right) ^{2}}{\eta_{i}+\eta_{j}}\arrowvert\bra{\psi_{i}}{\cal K}\ket{\psi_{j}}\arrowvert^{2}.
\end{equation}
It should be noted that in the case of unitary dynamics, where ${\cal K}$ is a fixed Hermitian operator on the system $\rho$, QFI becomes independent of the estimated parameter $\theta$ as shown in \cite{Luo2004}. If $\rho_{AB}$ is a $2\otimes d$-bipartite quantum state, and if the dynamics of the first subsystem is controlled by the local phase shift transformation $e^{-i\theta{\cal K}_{A}}$, with ${\cal K}={\cal K}_{A}\otimes\openone_{B}$, The QFI with local Hamiltonians becomes
\begin{equation}
	\mathcal{F}\left(\rho,{\cal K}_{A}\right)={\rm Tr}\left(\rho {\cal K}_{A}^{2}\right)-\sum_{i \ne j} \frac{2\eta_{i}\eta_{j}}{\eta_{i}+\eta_{j}}\arrowvert\bra{\psi_{i}}{\cal K}_{A}\ket{\psi_{j}}\arrowvert^{2}. \label{eq8}
\end{equation}
Similar to LQU (\ref{LQU}), the local Fisher quantum information (LQFI) is defined as the worst-case QFI on all local Hamiltonians ${\cal K}_{A}$ acting on the subsystem $A$ \cite{Kim2018} as
\begin{equation}
	\mathcal{Q}_{F}\left(\rho\right) = \underset{{\cal K}_{A}}{\text{min}}\mathcal{F}\left( \rho,{\cal K}_{A}\right) \label{eq9}
\end{equation}
It is another analytically computable measure of discord-like correlations and has all the desirable features that any acceptable non-classical correlation quantifier should fulfill. The local Hamiltonian ${\cal K}_{A}$ can be taken as ${\cal K}_{A}=\vec{\sigma}.\vec{r}$ for minimizing the above equation (\ref{eq9}) for any qubit-qudit system, with $\arrowvert\vec{r} \arrowvert=1$ and $\vec{\sigma}=\left(\sigma_{1},\sigma_{2},\sigma_{3}\right)$. In this case, it is easy to prove that ${\rm Tr}\left(\rho {\cal K}_{A}^{2}\right)=1$ and the second term in equation (\ref{eq8}) can be formulated as 
\begin{align}
	\sum_{i \ne j} \frac{2\eta_{i}\eta_{j}}{\eta_{i} + \eta_{j}}&\arrowvert\bra{\psi_{i}}{\cal K}_{A}\ket{\psi_{j}}\arrowvert^{2}=\vec{r}^{\dagger}.{\cal M}.\vec{r},
\end{align}
with the matrix elements of the symmetric matrix ${\cal M}\equiv[{\cal M}_{kl}]$ are given by
\begin{equation}
{\cal M}_{kl}=\sum_{i\ne j} \frac{2\eta_{i}\eta_{j}}{\eta_{i} + \eta_{j}} \bra{\psi_{i}}\sigma_{k}\otimes \openone_{B}\ket{\psi_{j}}\bra{\psi_{j}}\sigma_{l}\otimes \openone_{B}\ket{\psi_{i}} \label{Mkl}.
\end{equation}
Applying the minimization process over all possible local observables, as we already made by LQU (\ref{lqu}), we obtain an explicit closed form of LQFI as follows
\begin{equation}
	{\cal Q}_{F}\left(\rho\right) =1-\xi_{\max}[{\cal M}],\label{13}
\end{equation}
where $\xi_{\max}[{\cal M}]$ denotes the largest eigenvalue of the symmetric matrix ${\cal M}$ defined by (\ref{Mkl}).
\subsection{Quantum coherence}
Quantum coherence (QC) is one of the latest correlations formally introduced \cite{Baumgratz2014,Streltsov2017,Levi2014}, although its concept has been mentioned since the origins of quantum mechanics. Actually, QC is the basis of entanglement, multi-particle interference and other types of non-classical correlations, and plays a central role in quantum information processing \cite{Shi2017,Olaya-Castro2008,SlaouiSalah2020,Mohamed2017}. Therefore, in order to explore the QC in the context of quantum resource theory, quantum information science proposes rigorous notions and techniques, and several quantifiers have been proposed in finite dimensional systems. Radhakrishnan and his co-authors \cite{Radhakrishnan2016} have proposed a new quantifier of quantum coherence, combining both entropic and metric properties and satisfying all known criteria for a QC quantifier. It is defined as the square root of the quantum Jensen-Shannon divergence ${\cal J}\left(\rho,\rho_{d}\right)$ as
 \begin{equation}
 {\cal Q}_{C}\left(\rho\right)=\sqrt{{\cal J}\left(\rho,\rho_{d}\right)}, \label{qc}
 \end{equation}
where the quantum Jensen-Shannon divergence, as a distance measure between the state $\rho$ and the closest incoherent state $\rho_{d}$, is given by
\begin{equation*}
	{\cal J}\left(\rho,\rho_{d}\right)= S\left( \frac{\rho + \rho_{d}}{2}\right)-\frac{1}{2}\left[S\left( \rho\right) +S\left( \rho_{d}\right)\right],
\end{equation*}
with $\rho_{d}$ is the diagonal part of quantum state $\rho$ in the computational basis and $S\left( \rho\right) =-{\rm Tr}\left(\rho\log_{2}\rho\right)$ is the von Neumann entropy.

\section{Theoretical Model}\label{Sec3}
\subsection{The field interacts with two non-interacting qubits}\label{mjc}
Many physical phenomena can be modeled by two-level systems interacting with a harmonic oscillator, including atoms with electromagnetic fields \cite{Jaynes1963,Allen1987}, nuclear spin interactions with magnetic fields \cite{Rabi1936,Rabi1937}, electrons with crystal lattice phonon modes \cite{Holstein1959}, superconducting LC circuits \cite{Johansson2006,Chiorescu2004}, superconducting qubits with nanomechanical resonators \cite{Schwab2005,Irish2003}, among others. The Rabi Hamiltonian controls the behavior of all these systems. Analytical solutions for its eigenvalues and eigenvectors are still lacking despite the fact that it has been the subject of much investigation since it was originally presented within the framework of nuclear magnetic spin resonance. When the qubits are nearly resonant with the oscillator and the coupling forces are much smaller than the frequencies of the oscillator (i.e., in the rotating wave approximation), it is possible to use the counter-rotation terms to get the Jaynes-Cummings model having the Hamiltonian \cite{Jaynes1963},
\begin{equation}
	\hat{H}=w\hat{a}^{\dagger}\hat{a}+\sum\limits_{j=A,B}\frac{w_{0}}{2}\sigma_{j}^{z}+\gamma\sum\limits_{j=A,B}\left(\hat{a}\sigma_{j}^{+}+\hat{a}^{\dagger}\sigma_{j}^{-}\right).
\end{equation} 
In this context, $\omega_{0}$ denotes the transition frequency that characterizes the two levels of the atoms. The parameter $\gamma$ represents the constant of coupling between the atoms and the fields. Additionally, $\omega$ corresponds to the angular frequency of the single-mode field, while $\hat{a}^{\dagger}$ and $\hat{a}$ symbolize the creation and annihilation operators for the cavity mode. The atomic transition operators $\sigma_{j}^{\pm}$ and inversion operator $\sigma_{j}^{z}$ are defined as
\begin{equation*}
	\sigma_{j}^{+}=\ket{0}_{j}\bra{1}, \quad \sigma_{j}^{-}=\ket{1}_{j}\bra{0}, \quad \sigma_{j}^{z}= \ket{0}_{j}\bra{0}-\ket{1}_{j}\bra{1},
\end{equation*}
with $\ket{0}_{j}$ and $\ket{1}_{j}$ representing respectively the ground and excited states of the $j$-th qubit. In fact, the interaction between the system and field influences how the whole system evolves. However, the energy transfer between the system and the field occurs from the perspective of the qubits, and this is defined by the backaction term $\hat{a}\sigma_{j}^{+}$ and the relaxation term $\hat{a}^{\dagger}\sigma_{j}^{-}$. we assume that the particles A and B are resonantly coupled to the single-mode cavity field and that they are initially prepared in Werner-like states \cite{Werner1989} as $\rho_{AB}=\varsigma\ket{\phi}\bra{\phi}+(1-\varsigma)/4\openone$ with $\ket{\phi}= \sin\alpha\ket{00}+\cos\alpha\ket{11}$, described in terms of the purity $\varsigma$ and mixing $\alpha$. However, the initial state of the cavity field is adjusted to the vacuum state, i.e. $\rho_{F}=\ket{0}\bra{0}$. The density matrix operator of the system evolves to $\varrho_{AB}\left(t\right)={\rm Tr}_{F}\left[{\cal U}\left(t\right)\rho_{ABF}{\cal U}^{\dagger}\left(t\right)\right]$, where the time evolution operator ${\cal U}\left(t\right) =\exp\left(-iHt\right)$. In the standard basis $\left\lbrace \ket{00}, \ket{01}, \ket{10}, \ket{11}\right\rbrace$, the density matrix $\varrho_{AB}\left(t\right)$ is given by 
\begin{align}
	\varrho_{AB}\left( t\right)=&u\ket{00}\bra{00}+y\left[ \ket{01}\bra{01}+\ket{10}\bra{10}\right]\notag\\
	&+z\left[\ket{01}\bra{10}+\ket{10}\bra{01}\right]+v\ket{11}\bra{11}\notag\\&+w\left[\ket{00}\bra{11}+\ket{11}\bra{00}\right],
	\label{rho1}
\end{align}
with the elements
\begin{align*}
	&u = \frac{1-\varsigma}{4}+\frac{\varsigma\sin^{2}\alpha}{9} \left( \cos\left( \delta t\right)+2\right)^2,\\
	&v = \frac{1-\varsigma}{4}+ \varsigma\cos^{2}\alpha +\frac{2\varsigma}{9}\left(\cos\left( \delta t\right) -1\right)^{2}\sin^{2}\alpha,\\
	&w = \frac{\varsigma\sin2\alpha}{6}\left( \cos\left( \delta t\right)  +2\right),\\
	&z = \frac{\varsigma}{6}\sin^{2}\left( \delta t\right) \sin^{2}\alpha,\hspace{1cm}y= \frac{1-\varsigma}{4}+z,
\end{align*}
where $\delta=\sqrt{6}\gamma$. We need to determine the components provided by Eq.(\ref{wij}) to determine the analytical formula of LQU. Following some simplifications, we arrive at
\begin{equation}
	{\cal W}_{11}= \frac{2\left( y+\sqrt{y^{2}-z^{2}}\right) \left( u+v+2\sqrt{uv-w^{2}}\right)+4zw+uv^{2}}{\sqrt{2\left( y+\sqrt{y^{2}-z^{2}}\right)\left( u+v+2\sqrt{uv-w^{2}}\right)}},
\end{equation}
\begin{equation}
	{\cal W}_{22}= \frac{2\left( y+\sqrt{y^{2}-z^{2}}\right) \left( u+v+2\sqrt{uv-w^{2}}\right)-4zw-uv^{2}}{\sqrt{2\left( y+\sqrt{y^{2}-z^{2}}\right)\left( u+v+2\sqrt{uv-w^{2}}\right)}},
\end{equation}
\begin{align}
	{\cal W}_{33}=&\frac{u^{2}-4w^{2}}{2\left( u+v+2\sqrt{uv-w^{2}}\right)}+\frac{v^{2}-4z^{2}}{4\left( y+\sqrt{y^{2}-z^{2}}\right)}\notag\\
	&+\frac{u+v}{4}+\sqrt{uv-w^{2}}+\sqrt{y^{2}-z^{2}}+y^{2},
\end{align}
Thus, it is simple to check that ${\cal W}_{11} \geq {\cal W}_{22}$, then LQU is given in term of
\begin{equation}
{\cal U}_{Q}\left(\varrho_{AB}\right) =1-\max\left\lbrace{\cal W}_{11},{\cal W}_{33}\right\rbrace.
\end{equation}
To derive the explicit expression for LQFI, we first compute the eigenvalues and eigenstates of the density matrix (\ref{rho1}). The eigenvalues are explicitly given by
\begin{align}
	&\eta_{1}=\frac{1-\varsigma}{4}, \hspace{1.5cm} \eta_{2}=\eta_{1}+ \frac{\varsigma \sin^{2}\alpha\sin^{2}\left( \delta t\right)}{3},\\
	&\eta_{3,4}=\frac{1}{2}\left[u+v\mp\frac{1}{3}\sqrt{9\left(u-v\right)^{2}+\left( 2+\cos\left(\delta t\right)\right)^{2}\varsigma^{2}\sin^{2}2\alpha}\right],
	\label{eigen1}
\end{align}
and the corresponding eigenstates are given by
\begin{align}
	&\ket{\psi_{1,2}} = \frac{1}{\sqrt{2}}\left[\ket{01}\mp\ket{10}\right],\notag\\& \ket{\psi_{3,4}} = \frac{1}{\sqrt{\chi_{\mp}^{2}+1}}\left[\chi_{\mp}\ket{00}+\ket{11}\right],
\end{align}
where the quantities $\chi_{\mp}$ are
\begin{equation}
\chi_{\mp}=\frac{3\left(u-v\right)\mp \sqrt{9\left(u-v\right)^{2}+16\varsigma^{2}\left( 2+\cos\delta t\right)^{2} \sin^{2}2\alpha}}{4\varsigma\left( 2+\cos\delta t\right)\sin2\alpha}.
\end{equation}
Using (\ref{Mkl}), the non-zero components of the symmetric matrix ${\cal M}$ are the diagonal elements given by
\begin{align}
	{\cal M}_{11}&= \frac{2\eta_{1}\eta_{3}\left(\chi_{-}-1\right) ^{2}}{\left( \eta_{1}+\eta_{3}\right) \left( \chi_{-}^{2}+1\right)}+\frac{2\eta_{1}\eta_{4}\left( \chi_{+}-1\right)^{2}}{\left(\eta_{1}+\eta_{4}\right)\left(\chi_{+}^{2}+1\right)}\notag\\
	&+\frac{2\eta_{2}\eta_{3}\left(\chi_{-}+1\right)^{2}}{\left(\eta_{2}+\eta_{3}\right) \left(\chi_{-}^{2}+1\right)}+\frac{2\eta_{2}\eta_{4}\left( \chi_{+}+1\right)^{2}}{\left( \eta_{2}+\eta_{4}\right)\left(\chi_{+}^{2}+1\right)},
\end{align}
\begin{align}
	{\cal M}_{22}&= \frac{2\eta_{1}\eta_{3}\left( \chi_{-}+1\right) ^{2}}{\left( \eta_{1}+\eta_{3}\right) \left( \chi_{-}^{2}+1\right)}+\frac{2\eta_{1}\eta_{4}\left( \chi_{+}+1\right)^{2}}{\left(\eta_{1}+\eta_{4}\right)\left(\chi_{+}^{2}+1\right)}\notag\\
	&+\frac{2\eta_{2}\eta_{3}\left( \chi_{-}-1\right) ^{2}}{\left( \eta_{2}+\eta_{3}\right) \left( \chi_{-}^{2}+1\right)}+\frac{2\eta_{2}\eta_{4}\left( \chi_{+}-1\right)^{2}}{\left( \eta_{2}+\eta_{4}\right) \left( \chi_{+}^{2}+1\right)},
\end{align}
\begin{equation}
	{\cal M}_{33} = \frac{2\eta_{3}\eta_{4}\left( \chi_{-}\chi_{+}-1\right)^{2}}{\left( \eta_{3}+\eta_{4}\right)\left(\chi_{-}^{2}+1\right) \left( \chi_{+}^2+1\right)}.
\end{equation}
Similar to LQU, it is simple to see that ${\cal M}_{11} \geq {\cal M}_{22}$ and then the LQFI writes
\begin{equation}
{\cal Q}_{F}\left(\varrho_{AB}\right) =1-\max\left\lbrace {\cal M}_{11},{\cal M}_{33}\right\rbrace.
\end{equation}
Subsequently, we determined the amount of quantum coherence existing in the system using the quantum Jensen–Shannon divergence. For the density matrix (\ref{rho1}), the analytical expression of the quantum Jensen-Shannon divergence takes the form
\begin{align*}
	{\cal J}\left(\varrho_{AB},\varrho_{d}\right)=\frac{1}{2\log\left( 2\right)}&\left[\sum_{i=1}^{4}\eta_{i}\log \eta_{i}+u\log u +v\log v\right. \notag\\&\left. -2\sum_{j=1}^{4}x_{j}\log x_{j}+2y\log y\right],  
\end{align*}
and therefore the quantum coherence is ${\cal Q}_{C}\left(\varrho_{AB}\right)=\sqrt{{\cal J}\left(\varrho_{AB},\varrho_{d}\right)}$, where $x_{i}$ are eigenvalues of the matrix $\left(\varrho_{AB}+\varrho_{d}\right)/2$.
\begin{widetext}
	
	\begin{figure}[hbtp]
		{{\begin{minipage}[b]{.33\linewidth}
					\centering
					\includegraphics[scale=0.215]{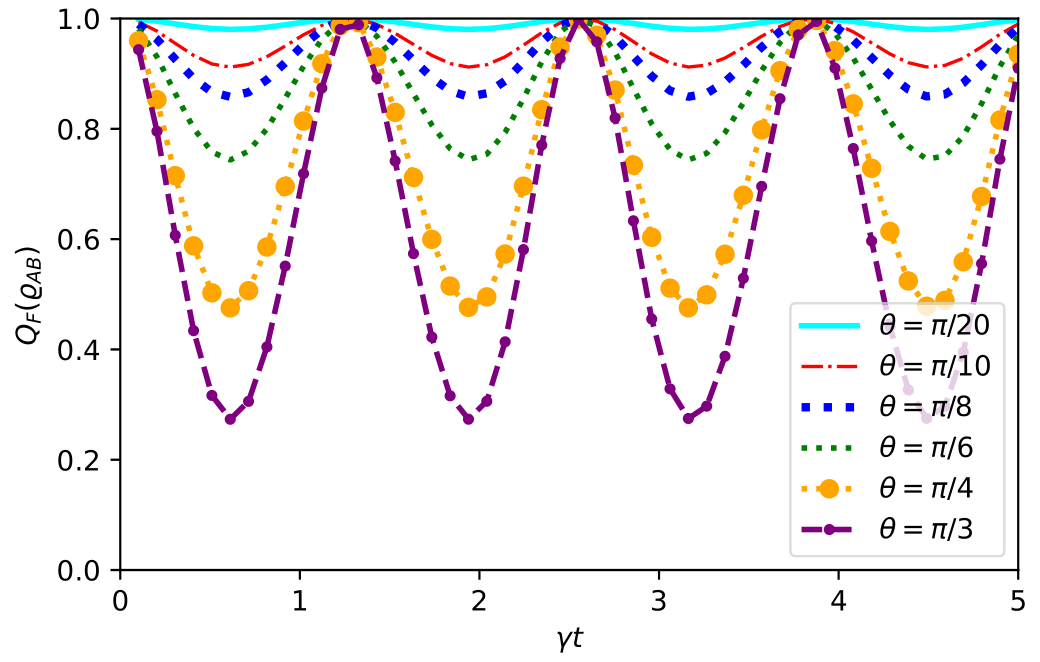} \vfill $\left(a\right)$
				\end{minipage}\hfill
				\begin{minipage}[b]{.33\linewidth}
					\centering
					\includegraphics[scale=0.215]{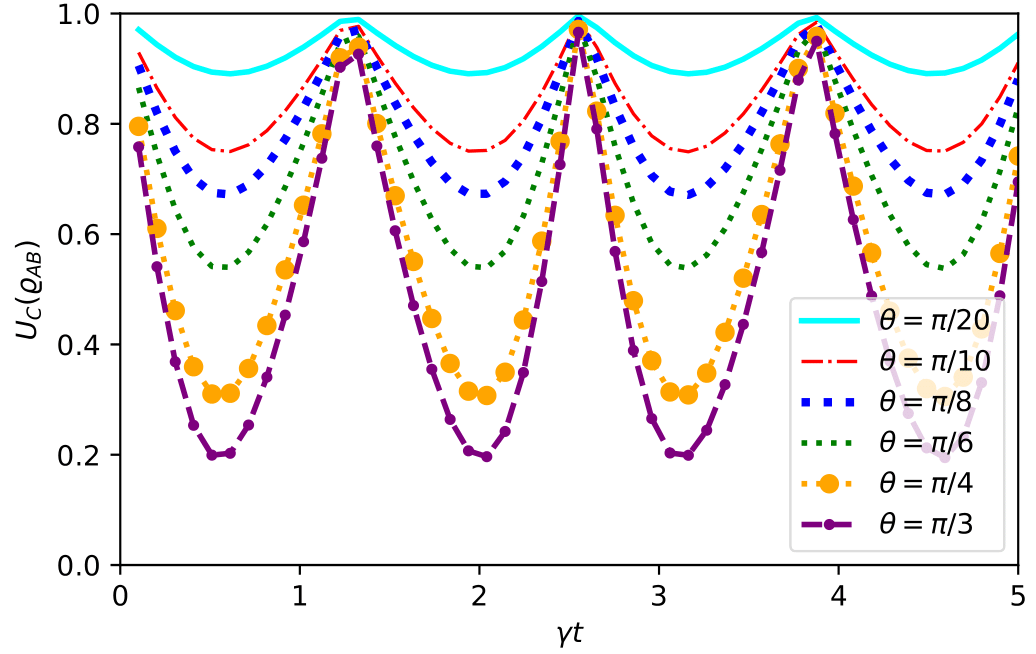} \vfill  $\left(b\right)$
				\end{minipage}\hfill
				\begin{minipage}[b]{.33\linewidth}
					\centering
					\includegraphics[scale=0.215]{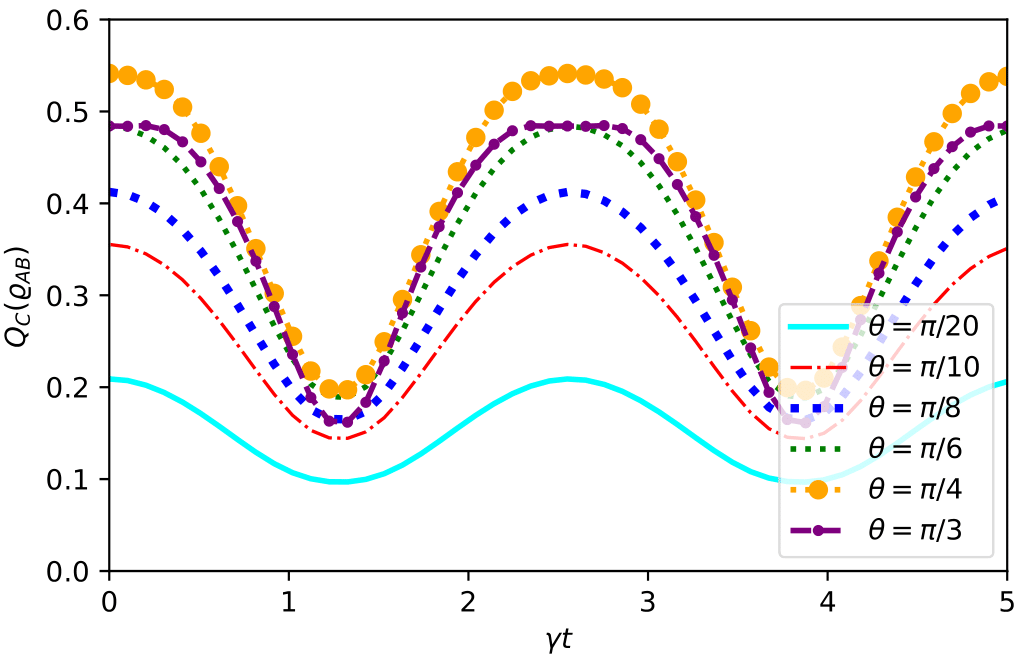} \vfill $\left(c\right)$
		\end{minipage}}}\\
		{{\begin{minipage}[b]{.33\linewidth}
				\centering
				\includegraphics[scale=0.215]{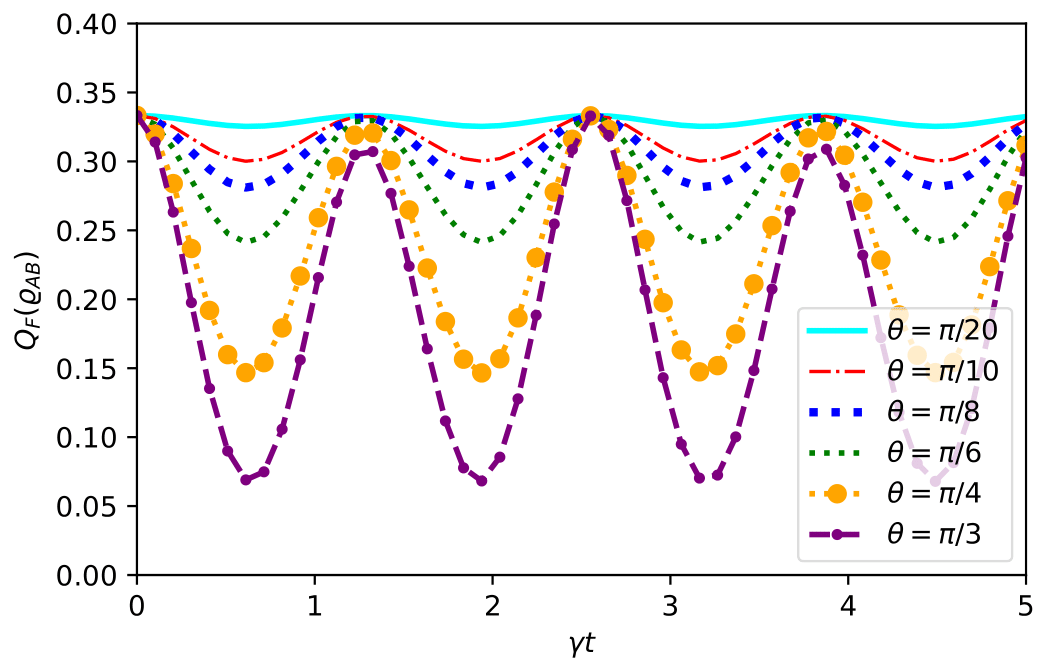} \vfill $\left(d\right)$
			\end{minipage}\hfill
			\begin{minipage}[b]{.33\linewidth}
				\centering
				\includegraphics[scale=0.215]{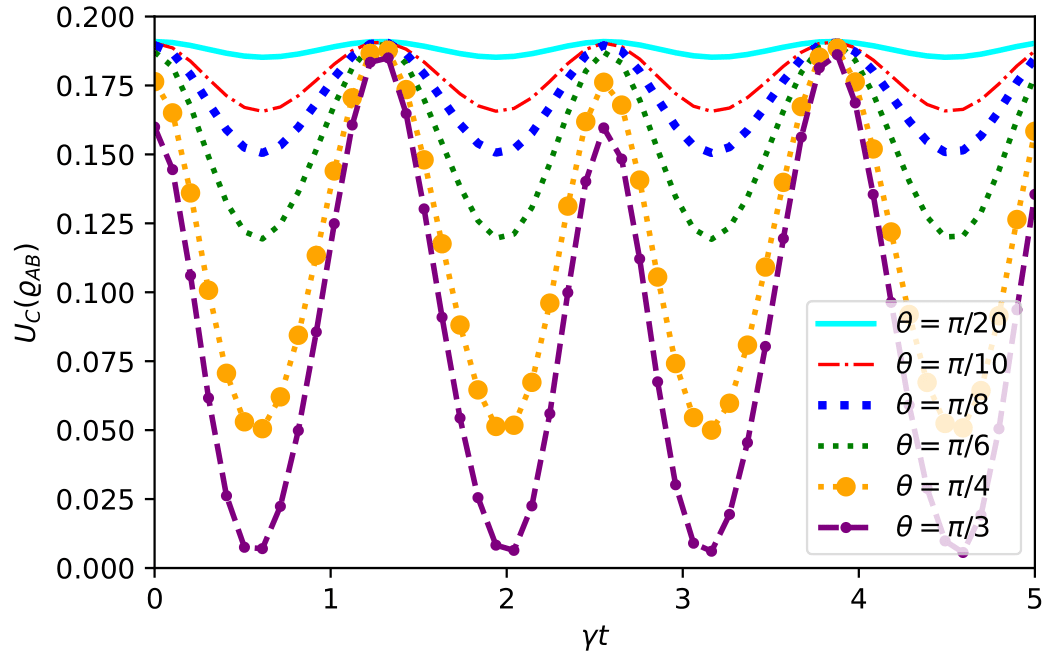} \vfill  $\left(e\right)$
			\end{minipage}\hfill
			\begin{minipage}[b]{.33\linewidth}
				\centering
				\includegraphics[scale=0.215]{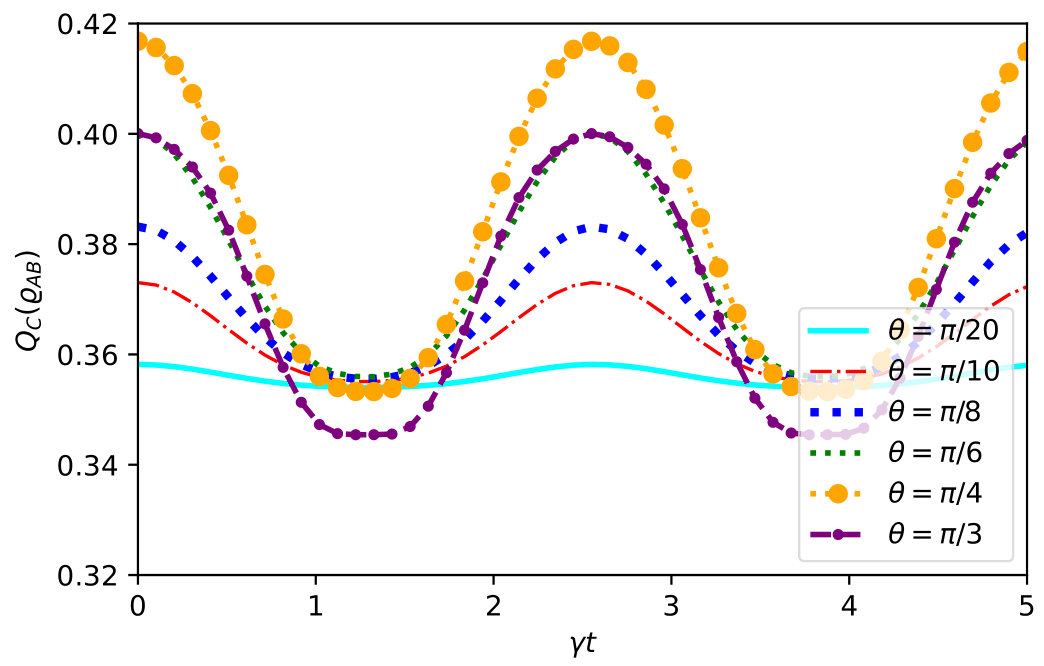} \vfill $\left(f\right)$
	\end{minipage}}}
		\caption{Time evolutions of local quantum uncertainty, quantum coherence, and local quantum Fisher information versus the coupling $\gamma t$ for different value of $\alpha$. The top row corresponds to $\varsigma=1$ and the bottom row corresponds to $\varsigma=0.5$.}\label{Fig1}
	\end{figure}
\end{widetext}

We can observe that the parameters of the initial state, purity $\varsigma$ and mixing $\alpha$ control the evolution of local quantum Fisher information, local quantum uncertainty and quantum coherence in the Jaynes-Cumming model. At $\varsigma=1$, the behavior of the LQFI is observed to be periodic, where it initially takes its maximum value and then decreases over time to its minimum value before rising again. This behavior repeats itself while maintaining the same amplitude and expansion. The control of the $\alpha$ value contributes to the minimization of quantum correlation collapses, whereby a smaller parameter value leads to fewer collapses, and vice versa, as depicted in Fig.\ref{Fig1}(a). Similarly, LQU exhibits the same periodic behavior in terms of both behavior and quantity, as illustrated in Fig.\ref{Fig1}(b).\par

On the other hand, QC behaves differently than LQFI and LQU, starting from the lowest value and gradually increasing to its maximum value. Raising the value of $\alpha$ also contributes to an increase in the amount of quantum coherence, opposite to LQU, and LQFI (see Fig.\ref{Fig1}). If we reduce the purity value to $0.5$, the same behavior is maintained, but the quantity is different $\mathcal{Q_{C}}|_{\varsigma=1}>\mathcal{Q_{C}}|_{\varsigma=0.5}$ the same for LQFI and LQU (see fig.\ref{Fig1}(d),(e) and (f)). The degradation of quantum correlations can be attributed to the decoherence of qubits resulting from their interaction with the environment. As time progresses, the two qubits exchange energy with the environment, which leads to a loss of coherence and an increase in the number of classical correlations. Moreover, the eigenstate of the system also contributes to the degradation of quantum correlations.
\subsection{Non-Markovian dephasing model for a double two-qubit system}\label{nmm} 
In this part, we explore the dynamics of the aforementioned quantum criteria in the case of the system-environment interactions with memory, but without taking into account the Born-Markov approximation which assumes that the system-environment correlation time is infinitely short in order to neglect the memory effects \cite{Daffer2004,Pinto2013}. Firstly, we focus on a colored noise dephasing model where its dissipative dynamics is described by the following master equation
\begin{equation}
	\dot{\rho}\left( t\right) = K L\rho, \label{mitraise}
\end{equation}
$K$ stands for a time-dependent integral operator whose action on the system is given by $K\varphi=\int_{0}^{t} k\left( t-t^{,}\right)\varphi\left( t^{,}\right)dt^{,}$, $L$ is a Lindblad superoperator, $\rho$ is the density operator of the principal system and $k\left(t-t^{,}\right)$ represents the kernel function allowing to determine the type of memory in a specific environment. Indeed, even in the absence of the integral operator $K$ in Eq.(\ref{mitraise}), we typically obtain the master equation with Markovian approximation. If a two-level quantum system interacts with a reservoir that exhibits random telegraph signal noise characteristics, then this type of master equation could arise. The time-dependent Hamiltonian
\begin{equation}
	H\left( t\right) =\hbar \sum\limits_{i=1}^{3} \Gamma_{i}\left( t\right)\sigma_{i},
\end{equation}
$\Gamma_{i}\left(t\right)$ are independent random variables complying with the characteristics of a random telegraph signal and $\sigma_{i}$ are the standard Pauli matrices. The random variables can be specifically stated as $\Gamma_{i}\left( t\right) =a_{i}{n_{i}\left( t\right)}$, with $n_{i}\left( t\right)$ has a Poisson distribution with a mean of $t/2\tau_{i}$ and $a_{i}$ is an independent random variable having values $\pm a_{i}$. The random telegraph signal is described by a broadly defined stationary stochastic process with zero mean. This concept can be used for any two-level quantum system that interacts with a random telegraph signal noise. This can be applied to a two-level atom exposed to a fluctuating laser field with random jump-like phase noise \cite{Kampen1992,Eberly1984}. The parameters $a_{i}$ determine the degree of coupling of the system with the external impact. The inverse of $\tau_{i}$ is the flipping or fluctuation rate. Applying the von Neumann equation of motion $\dot{\rho}\left( t\right)=-i\sum_{i} \Gamma\left( t\right)\left[ \sigma_{i},\rho\right]$, the matrix density of a two-level system can be written as
\begin{equation}
	\rho\left( t\right) =\rho\left( 0\right) -i\int_{0}^{t} dk \sum_{i} \Gamma\left(k\right)\left[ \sigma_{i},\rho\left( k\right)\right].\label{eq31}
\end{equation}
Substituting Eq.(\ref{eq31}) into the von Neumann equation and carrying out the stochastic averages, we find ourselves with
\begin{equation}
	\dot{\rho}\left( t\right)=-\int_{0}^{t} \sum\limits_{k} \exp\left(-\frac{\mid t-t^{,}\mid}{\tau_{k}}\right) a_{k}^{2} \left[ \sigma_{k},\left[ \sigma_{k},\rho\left( t^{,}\right) \right] \right] dt^{,}.\label{rhooo}
\end{equation}
The memory kernel derived from the correlation functions of the random telegraph signal takes the form
\begin{equation}
	<\Gamma_{j}\left( t\right) \Gamma_{k}\left( t^{,}\right)>=a_{k}^{2} \exp\left(- \frac{\arrowvert t-t^{,}\arrowvert}{\tau_{k}}\right) \delta_{jk}.
\end{equation}
The model proposed above generates a non-Markovian time evolution when one takes into account some recently developed measures of non-markovianity \cite{Breuer2009,Luo2012}. The dynamic evolution produced by Eq.(\ref{rhooo}) appears to be entirely positive if two of the $a_{k}$ are zero. This would be analogous to a physical scenario in which noise only comes from one direction. Notably, the system dynamics is those of a dephasing with colored noise if $a_{1} = a_{2} = 0$ and $a_{3} = a$, and in such case, the Kraus operators which is related to the dynamics of two-level system are \cite{Daffer2004}
\begin{equation*}
	F_{1}\left(t \right)=\sqrt{\frac{1+\Lambda\left(\xi,\kappa\right)}{2}}\openone_{2\times2},\hspace{0.5cm}
	F_{2}\left(t\right)=\sqrt{\frac{1-\Lambda\left(\xi,\kappa\right)}{2}}\sigma_{3}.
\end{equation*}
where
\begin{align}
	&\Lambda\left(\xi,\kappa\right)=e^{-\xi}\left(\cos(\kappa\xi)+\sin(\kappa\xi)/\kappa\right), \notag\\& \kappa=\sqrt{\left(4a\tau\right) ^{2}-1},\hspace{1cm}\xi=\frac{t}{2\tau},
\end{align}
and the Kraus operators $F_{i}\left(t\right)$ satisfying the normalization condition $\sum_{i}F_{i}^{\dagger}\left(t\right)F_{i}\left( t\right) =\openone$. For bipartite systems, the time evolution of an initial density operator is described as follows
\begin{equation}
	\rho_{AB}\left(t\right)=\sum_{i,j}\left(F_{i}^{A}\otimes F_{i}^{B}\right) \rho_{AB}\left(0\right)\left(F_{i}^{A}\otimes F_{i}^{B}\right)^{\dagger} 
\end{equation}
where the operators $F_{i}^{A}$ and $F_{i}^{B}$ respectively act on the first and second qubits, while $\rho_{AB}\left(0\right)$ denotes the initial state of the two-qubit system. In this part, we select the same initial state as in the first model, and after damping, the density operator takes the form
\begin{align}
	\rho_{AB}\left(t\right)&=A\ket{00}\bra{00}+D\left( \ket{01}\bra{01}+\ket{10}\bra{10}\right)\notag\\
	&+C\left(\ket{00}\bra{11}+\ket{11}\bra{00}\right)+B\ket{11}\bra{11},\label{rho2}
\end{align}
with the entries notations are adopted as
\begin{align*}
	&A=\frac{1-\varsigma}{4}+\varsigma\sin^{2}\alpha, \quad B=\frac{1-\varsigma}{4}+\varsigma\cos^{2}\alpha,\\
	 &C=\frac{\varsigma}{2}\Lambda\left(\xi,\kappa\right)\sin2\alpha, \hspace{1cm} D=\frac{1-\varsigma}{4}.
\end{align*}
The eigenvalues of the density operator (\ref{rho2}) are
\begin{align}
	&\lambda_{1,2}=\frac{1-\varsigma}{4},\notag\\
	&\lambda_{3,4}=\frac{1}{4}\left( 1+\varsigma\mp\sqrt{2\varsigma^{2}\left( 1+\cos4\alpha\right) +4\Lambda\left(\xi,\kappa\right)^{2}\sin^{2}2\alpha}\right),
	\label{lamb}
\end{align}
and the corresponding eigenstates are given by
\begin{align}
	&\ket{\psi_{1}}=\ket{10}, \quad \ket{\psi_{2}}=\ket{01},\notag\\
	&\ket{\psi_{3,4}}=\frac{1}{\sqrt{\varepsilon_{\mp}^{2}+1}}\left(\varepsilon _{\mp}\ket{00}+\ket{11}\right),
\end{align}
with
\begin{equation}
\varepsilon_{\mp}=\frac{1}{\sin\left(2\alpha\right)\Lambda\left(\xi,\kappa\right)}\left(-\cos2\alpha\mp\varsigma\sqrt{\Lambda\left(\xi,\kappa\right)^{2}\sin^{2}2\alpha+\cos^{2}2\alpha}\right).
\end{equation}
Based on the above formalism, the elements of the $3×3$ symmetric matrix ${\cal M}$ are calculated analytically from Eq.(\ref{Mkl}) and found to be as follows
\begin{align}
&{\cal M}_{11}={\cal M}_{22}=4\left(\frac{\lambda_{1}\lambda_{3}}{\lambda_{1}+\lambda_{3}}+\frac{\lambda_{1}\lambda_{4}}{\lambda_{1}+\lambda_{4}}\right),\notag\\&
	{\cal M}_{33}= \frac{4\lambda_{3}\lambda_{4}}{\lambda_{3}+\lambda_{4}}\frac{\left( \varepsilon_{+}\varepsilon_{-}-1\right)^{2}}{\left( \varepsilon_{+}^{2}+1\right)\left( \varepsilon_{-}^{2}+1\right)},
\end{align}
and to obtain the analytical form of the local quantum Fisher information (\ref{13}), it is necessary to treat separately two cases, i.e. ${\cal M}_{11}\geq{\cal M}_{33}$ or ${\cal M}_{33}\geq{\cal M}_{22}$ with which
\begin{equation}
	{\cal Q}_{F}\left(\rho_{AB}\right) =1-\max\left\lbrace {\cal M}_{11},{\cal M}_{33}\right\rbrace.
\end{equation}
Likewise, the explicit expressions of the matrix elements ${\cal W}_{ij}$ (\ref{wij}) can be found as 
\begin{align}
	&{\cal W}_{11}={\cal W}_{22}=\frac{\Theta^{2}+\varsigma^{2}\Lambda\left(\xi,\kappa\right)^{2}\sin^{2}(2\alpha)}{4\sqrt{\Theta}},\notag\\&
	{\cal W}_{33}=\frac{\left(3-\varsigma+2\sqrt{\Sigma}\right)^{2}+8\varsigma^{2}\Lambda\left(\xi,\kappa\right)^{2}\sin^{2}(2\alpha)}{8\left(3-\varsigma+2\sqrt{\Sigma}\right)},
\end{align}
where
	\begin{align}
	\Theta=&2\left(\varsigma-1\right)\sqrt{1+\left(2-3\varsigma\right)\varsigma-4\varsigma^2\left(\Lambda\left(\xi,\kappa\right)^{2}-1\right)\sin^{2}(2\alpha)}\notag\\&+2\left(\varsigma^{2}-1\right),
\end{align}
\begin{align}
	\Sigma=1+\left(2-3\varsigma\right)\varsigma-4\varsigma^{2}\left(\Lambda\left(\xi,\kappa\right)^{2}-1\right)\sin^{2}(2\alpha).
\end{align}
Then the local quantum uncertainty is derived by
\begin{equation}
	{\cal U}_{Q}\left(\rho_{AB}\right)=1-\max\left\lbrace{\cal W}_{11},{\cal W}_{33}\right\rbrace.
\end{equation}
On the other side, we determined the analytic evolution of the quantum coherence after the damping by means of Eq.(\ref{qc}). Following a straightforward calculation, the analytical expression of the quantum coherence for the non-Markovian dephasing model described by the density matrix (\ref{rho2}) is
\begin{align*}
{\cal Q}_{C}\left(\rho_{AB}\right) &=\left[\frac{1}{16\log 16}\left(\varsigma-1 -2\varsigma\cos^{2}\alpha +\left(1+\varsigma\right)\log240\right)\right. \\
	&\left. +\left(\frac{1+\varsigma\mp2\varsigma\cos2\alpha}{2\log 16}\right)\log\left( \frac{1+\varsigma\mp2\varsigma\cos2\alpha}{4}\right)\right. \\
	&\left. - \frac{\varsigma\sin^{2}\alpha}{2}- \left(\frac{2+2\varsigma\mp\sqrt{\varpi}}{2\log 16}\right)\log \left(\frac{2+2\varsigma\mp\sqrt{\varpi}}{4}\right)\right. \\
	&\left. + \left( \frac{ 1+\varsigma\mp\sqrt{\Delta}}{\log 16}\right)\log \left(\frac{1+\varsigma\mp\sqrt{\Delta}}{4} \right)\right]^{\frac{1}{2}},
\end{align*}
where
\begin{align}
&\varpi=2\varsigma^{2}\left(4+\Lambda\left(\xi,\kappa\right)^{2}+\left(4-\Lambda\left(\xi,\kappa\right)^{2}\right)\cos(4\alpha)\right),\notag\\&\Delta=2\varsigma^{2}\left( 1+\Lambda\left(\xi,\kappa\right)^{2}+\left(1-\Lambda\left(\xi,\kappa\right)^{2}\right)\cos(4\alpha)\right).
\end{align}

\begin{widetext}
	
	\begin{figure}[hbtp]
		{{\begin{minipage}[b]{.33\linewidth}
					\centering
					\includegraphics[scale=0.29]{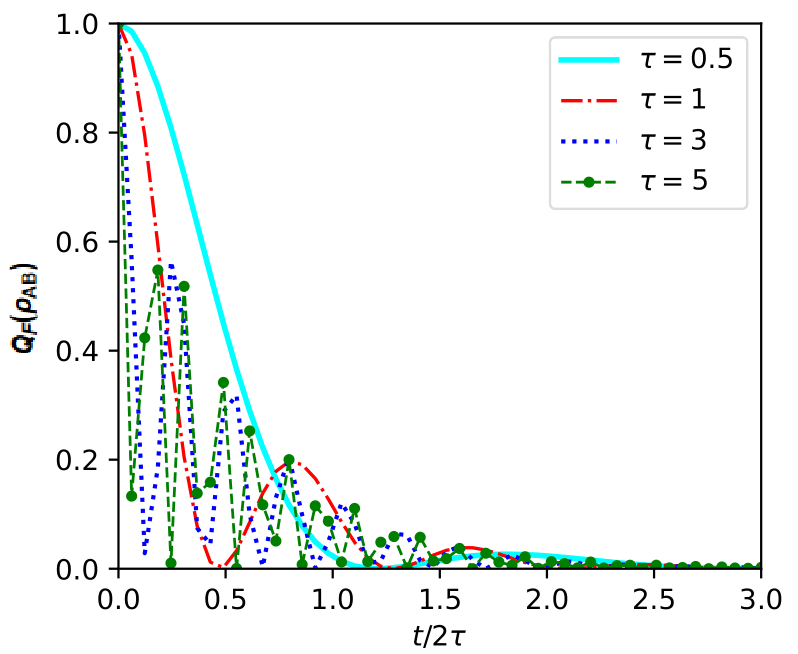} \vfill $\left(a\right)$
				\end{minipage}\hfill
				\begin{minipage}[b]{.33\linewidth}
					\centering
					\includegraphics[scale=0.29]{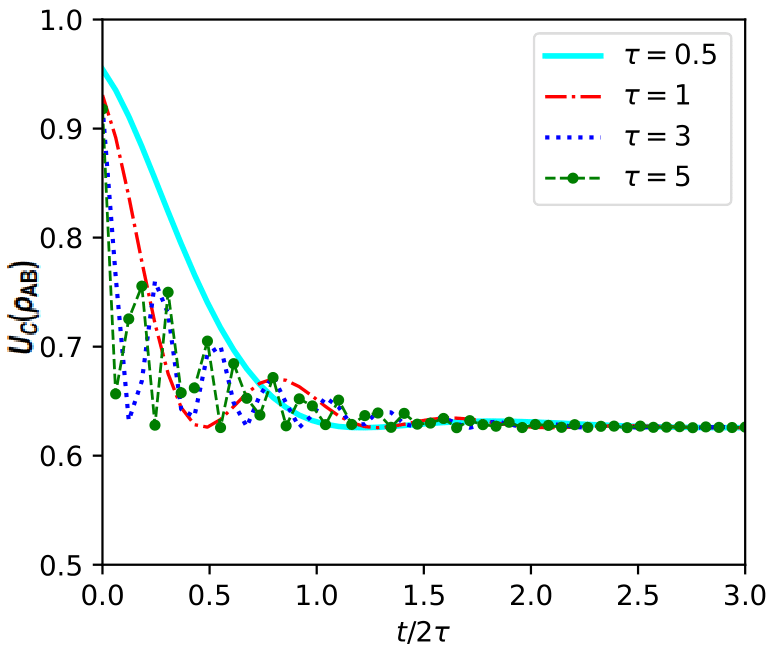} \vfill  $\left(b\right)$
				\end{minipage}\hfill
				\begin{minipage}[b]{.33\linewidth}
					\centering
					\includegraphics[scale=0.29]{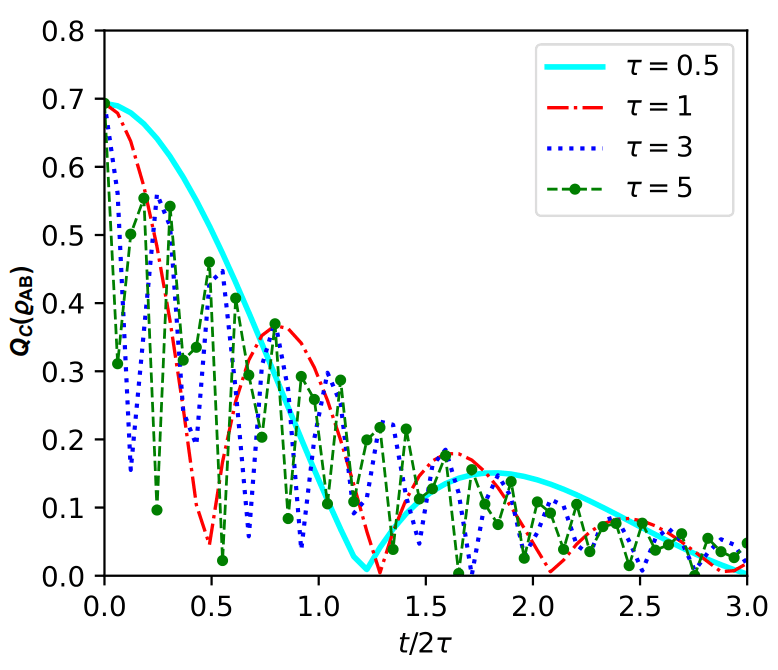} \vfill $\left(c\right)$
		\end{minipage}}}\\
		{{\begin{minipage}[b]{.33\linewidth}
					\centering
					\includegraphics[scale=0.29]{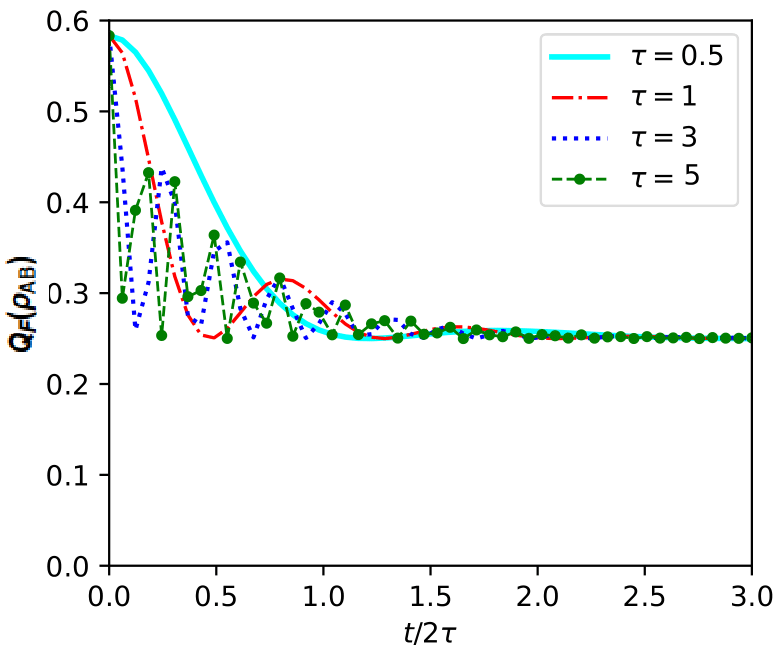} \vfill $\left(d\right)$
				\end{minipage}\hfill
				\begin{minipage}[b]{.33\linewidth}
					\centering
					\includegraphics[scale=0.29]{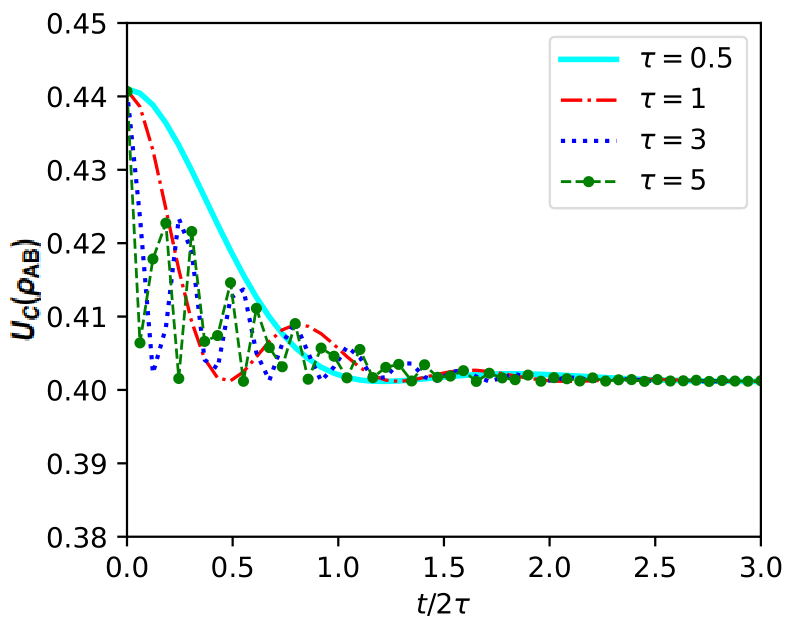} \vfill  $\left(e\right)$
				\end{minipage}\hfill
				\begin{minipage}[b]{.33\linewidth}
					\centering
					\includegraphics[scale=0.29]{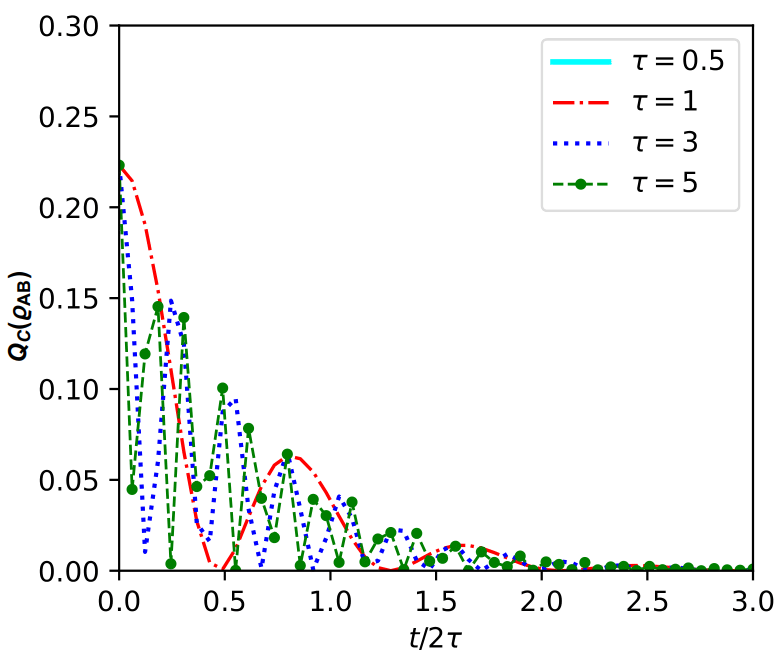} \vfill $\left(f\right)$
		\end{minipage}}}
		\caption{Dynamics of LQFI, LQU and QC as a function of $t/2\tau$ for different values of the degree of non-Markovianity $\tau$. Panels ($a$)-($b$)-($c$) corresponds to the pure initial state with $\varsigma=1$, while panels ($d$)-($e$)-($f$) correspond to the mixed initial state with $\varsigma=0.5$. Parameters $a$ and $\alpha$ are fixed as $a=1$ and $\alpha=\pi/4$.}\label{Fig2}
	\end{figure}
\end{widetext}

To investigate the dynamic behavior of local quantum uncertainty and quantum coherence, as well as local quantum Fisher information, and explore how both markovian and non-markovian environments influence their evolution, we depicted their dynamics in Fig.(\ref{Fig2}) for a two-qubit system interacting with independent colored dephasing reservoirs. The parameter $t/2\tau$ was employed as the variable. It is worth noting that in Fig.(\ref{Fig2}), where $\varsigma=1$ (i.e. pure initial state) is fixed for panels ($a$-$b$-$c$) and $\varsigma=0.5$ (i.e. mixed initial state) is fixed for panels ($d$-$e$-$f$), the values $\tau=0.5$ and $\tau=1$ represent the markovian case, while $\tau=3$ and $\tau=5$ correspond to the non-markovian regime.\par

For the non-markovian regime (i.e., $\tau\geq2$), the dynamics of non-classical correlations (LQU and LQFI) and quantum coherence exhibit a phenomenon of death and revival, where their amplitudes undergo continuous damping. This behavior arises from the effects of environmental memory, which facilitate the back-flow of information. In contrast, for the markovian regime (i.e., $\tau<2$), the QC, LQU and LQFI asymptotically approach zero as time increases, with a little revival. This trend can be attributed to the rapid outflow of quantum information from the system to the environment, facilitated by weak system-environment coupling and the transfer of information without memory. Furthermore, a decrease in the degree of non-markovianity $\tau$ leads to a reduction in the frequency and delay of oscillations in the correlation function, as well as a decrease in its amplitude. Additionally, decreasing the purity results in a decrease in the amount of quantum criteria. Moreover, as depicted in Fig.(\ref{Fig2}), all aspects of quantumness (LQU, LQFI, and QC) evolve simultaneously and at the same rate. Also, in the non-Markovian dephasing model, the quantumness is higher when the initial state is pure compared to when it is mixed. This observation aligns with previous findings in the Jaynes-Cummings model.\par

As time progresses in the non-Markovian dephasing model, the correlation amplitudes of LQFI and LQU decrease due to the strong influence of the environment. However, it is noteworthy that the qualitative agreement between the LQFI and LQU functions remains intact. Notably, the LQFI exhibits higher correlation than the LQU, implying that the dynamics of LQFI and LQU align with the principles of quantum-Fisher and Wigner-Yanase skew information theories \cite{SlaouiB2019}. Specifically, the amplitudes of LQFI are consistently much larger than those of the LQU. Besides, QC exhibits greater robustness compared to LQFI and LQU. In fact, while the non-classical correlations vanished, the QC remained non-zero. This clear distinction indicates that under a colored noise dephasing model, quantum coherence is more resilient and robust in comparison to LQFI and LQU.

\section{Quantum teleportation scheme}\label{Sec4}
Quantum teleportation is a captivating phenomenon in the realm of quantum physics, challenging our classical understanding \cite{Bennett1993,Bowen2001,Ji-Gang2017}. In contrast to the instantaneous matter transfer of science fiction, quantum teleportation revolves around exchanging quantum information between distant particles. This extraordinary feat is achievable through quantum entanglement. Its significance lies in its vital role in quantum computing, secure quantum communication, and the profound exploration of quantum reality's essence. Despite its current limitation in transporting physical objects or individuals across long distances, quantum teleportation paves the way for thrilling prospects in the realms of quantum physics and advanced quantum technologies. To achieve quantum teleportation, cientists exploit the fascinating phenomenon of entanglement between two particles, called entangled particles. The idea is to have a third particle, which we want to teleport, and a fourth particle entangled with the third and located where we want to "teleport" the information \cite{KirdiA2023,Ikken2023,Kirdi22023}. By performing complex quantum measurements on the two entangled particles and the particle to be teleported, we can transfer the quantum state of the original particle to the destination particle. This happens without any physical transport of matter, as only the quantum state is transferred.

\begin{figure}[h]
	\centering
	\includegraphics[width=1\linewidth]{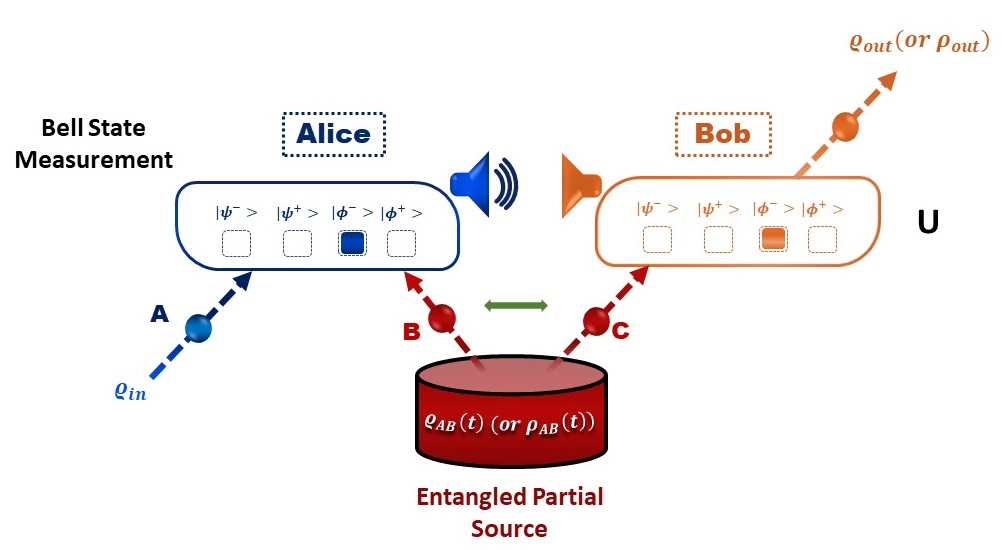}
	\caption{Quantum teleportation protocol for the transmission of quantum information from Alice to Bob through the entanglement of qubits: Alice possesses a qubit $A$ in an initial state $\rho_{in}$ that she wishes to teleport to Bob. To facilitate the process, Alice and Bob share an ancillary entangled pair of qubits ($B$ and $C$), generated by an Entangled partial source. Alice performs a joint Bell state measurement on qubit $A$ and qubit $B$, projecting them onto one of the four orthogonal entangled states known as the Bell states ($\ket{\psi^{\mp}}$, $\ket{\phi^{\mp}}$). The measurement result is then sent to Bob through classical communication. Depending on the outcome, Bob either keeps qubit $C$ (case $\ket{\phi^{-}}$) unchanged (case $\ket{\psi^{-}}$) or applies a unitary transformation ($U$) to qubit $C$ (cases $\ket{\psi^{+}}$ and $\ket{\phi^{+}}$). This transformation ensures that the output state $\varrho_{out}$ (or $\rho_{out}$) of qubit $C$ is identical to the input state $\rho_{in}$ of the original qubit $A$, achieving successful quantum teleportation.}
	\label{QT}
\end{figure}

In this scenario, we employ a teleportation protocol proposed by Cola and Paris \cite{Cola2005} to transfer a two-qubit system using only a single quantum channel. The protocol utilizes a single entangled two-qubit pair, the transmission of three classical bits, and an additional qubit introduced by the receiver (see Fig.(\ref{QT})). Standard teleportation can be regarded as a general depolarizing channel \cite{Brown2001}. In the two models discussed earlier (J-C and non-Markovian dephasing model), we are examining how the initial state impacts the efficiency of teleportation. In this context, Alice and Bob share a partially entangled state described by Eq.(\ref{rho1}) (for the JC model) and Eq.(\ref{rho2}) (for the non-Markovian dephasing model). Alice's unknown state is given in the form $\rho_{in}=\left|\psi_{in} \right\rangle\left\langle \psi_{in}\right|$, where 
\begin{equation}
	\ket{\psi_{in}}=\cos\left( \frac{\theta}{2}\right)\ket{10}+e^{-i\varphi}\sin\left( \frac{\theta}{2}\right)\ket{01},\label{eq47}
\end{equation}
with $0\leq\theta \leq \pi$ and $0\leq\varphi \leq 2\pi$. Here, $\theta$ describes an arbitrary state and $\varphi$ is the corresponding phase of this state. At the concluding step of the teleportation protocol, Bob will receive the state
\begin{equation}
\rho_{out}=\sum\limits_{kl=\left\lbrace 0,1,2,3\right\rbrace}p_{kl}\left( \sigma_{k} \otimes\sigma_{l}\right) \rho_{in}\left(\sigma_{k} \otimes\sigma_{l}\right),
\end{equation}
where $p_{kl}={\rm Tr}\left[K^{k}\rho_{ch}\right]{\rm Tr}\left[K^{l}\rho_{ch}\right]$, $\sigma_{n}$ ($n=k,l$) are the Pauli operators, $\sum_{kl}p_{kl}=1$ and $\rho_{ch}$ and A stands for the resource state linking Alice and Bob. Here, the different projection operators $K^{l}$ are defined as
\begin{align}
&K^{0,3}=\ket{\psi^{\mp}}\bra{\psi^{\mp}},\hspace{0.5cm}K^{1,2}=\ket{\phi^{\mp}}\bra{\phi^{\mp}},\notag\\
	&\ket{\phi^{\pm}}=\frac{1}{\sqrt{2}}\left(\ket{00}\pm \ket{11}\right),\hspace{0.5cm}\ket{\psi^{\pm}}=\frac{1}{\sqrt{2}}\left(\ket{01}\pm \ket{10}\right).\label{eq49}
\end{align}
The Bell states mentioned above (\ref{eq49}) exhibit maximum entanglement, implying an inseparable connection between their component qubits. Such states have been utilized as quantum channels to experimentally transmit an arbitrary state from one distant location to another. Nonetheless, in the two models under consideration, it is evident that the states may not be maximally entangled due to the presence of decoherence arising from the interaction between the system and its environment. However, our findings demonstrate that by manipulating the parameters within the reduced density operator, it becomes possible to attain a significant level of quantum correlations. To evaluate the effectiveness of the teleportation process, it is valuable to examine the fidelity between $\rho_{in}$ and $\rho_{out}$, which measures how close the final state is to the initial state. When the input state is a pure state, fidelity serves as a beneficial metric for assessing the teleportation performance of a quantum channel quantifier \cite{Jozsa1994}. It is defined as
\begin{equation}
F\left(\rho_{in},\rho_{out}\right)=\left\lbrace {\rm Tr}\left[ \sqrt{\sqrt{\rho_{in}}\rho_{out}\sqrt{\rho_{in}}}\right] \right\rbrace^{2}.\label{eq50}
\end{equation}
As the specific state to be teleported is generally unknown, it proves more beneficial to compute the average fidelity, which represents the mean value of fidelity for the four outcomes resulting from Bell state measurements \cite{Bowdrey2002}. This average fidelity is expressed as
\begin{equation}
F_{av}\left(\rho_{in},\rho_{out}\right)=\frac{1}{4\pi}\int_{0}^{2\pi}d\varphi\int_{0}^{\pi}F\left(\rho_{in},\rho_{out}\right)\sin\theta d\theta,\label{Fav}
\end{equation}
where $4\pi$ is the solid angle. If Alice and Bob share the quantum state $\varrho_{AB}\left( t\right)$ that describes the Jaynes-Cummings model, i.e. $\rho_{ch}=\varrho_{AB}\left(t\right)$, the resulting output state can be expressed as follows:
\begin{align}
\varrho_{out}\left(t\right)=&\kappa\left( \ket{00}\bra{00}+\ket{11}\bra{11}\right)+\chi \ket{01}\bra{01}\notag\\
&+\vartheta\left( \ket{00}\bra{11}+\ket{11}\bra{00}\right)+\zeta\ket{10}\bra{10}\notag\\
&+\Delta\ket{01}\bra{10}+\Theta\ket{10}\bra{01},
\label{rhoout1}
\end{align}
with their entries are given by
\begin{align}
&\kappa=2y\left( u+v\right),\hspace{1cm}\vartheta=4zw\cos\varphi \sin\theta,\notag\\
&\chi=\left( 4y^{2}+\left(u+v\right)^{2}\right) \sin^{2}\left( \theta/2\right),\notag\\
&\zeta=\left( 4y^{2}+\left(u+v\right)^{2}\right) \cos^{2}\left( \theta/2\right),\notag\\
&\Delta=2\varsigma e^{-i\varphi}\left(z^{2}+w^{2}e^{2i\varphi}\right)\sin\theta,\notag\\
&\Theta=2\varsigma e^{-i\varphi}\left(w^{2}+z^{2}e^{2i\varphi}\right)\sin\theta,
\end{align}
which are written according to the entries of the entangled resource state $\varrho_{AB}\left(t\right)$ (Eq.\ref{rho1}).
After some calculations, the fidelity of Bob's state takes the form
\begin{align}
F\left(\varrho_{out}\right)&=\frac{1}{188}\Bigg[\Big( 6+5\varsigma+\varsigma\cos2\alpha+2\varsigma\cos\Big(2\delta t\Big) \sin^{2}\alpha\Big)^{2}\times\notag\\
&\Big( 1+\cos^{2}\alpha\Big)+4\sin^{2}\theta\Big(4\varsigma^{2}\Big( 2+\cos\left(\delta t\right)  \Big) ^{2}\sin^{2}2\alpha\notag\\
&+\Big( 3-3\varsigma+2\varsigma\sin^{2} \left(\delta t\right) \sin^{2}\alpha\Big)^{2}\Big)\Bigg].\label{F1}
\end{align}
By inserting equation (\ref{F1}) into equation (\ref{Fav}), the result yields the average fidelity
\begin{align}
F_{av}\left(\varrho_{out}\right) &=\varsigma\Bigg[\frac{4}{9}\cos\left(\delta t\right) \sin^{2}2\alpha-\frac{5\varsigma}{64}\cos4\alpha+\notag\\
&\frac{1}{108}\Big( 6+19\varsigma+7\varsigma\cos2\alpha\Big )+\frac{\varsigma}{72}\cos\left( 4\delta t\right) \sin^{4}\alpha \Bigg]\notag\\
&+\frac{1}{4}+\frac{\varsigma}{144}\left(4+9\varsigma\right) \cos2\alpha+\frac{\varsigma}{576}\Big( 80+135\varsigma\Big)\notag\\&+\frac{\varsigma^{2}}{18}\sin^{2}\left(\delta t\right) \sin^{6}\alpha \cos 2\varphi.
\end{align}
If Alice and Bob share the state $\rho_{AB}\left(t\right)$ (eq.\ref{rho2}) with $\rho_{ch}=\rho_{AB}\left(t\right)$, which characterizes the non-Markovian dephasing model, then the resulting output state is given by
\begin{align}
	\rho_{out}\left(t\right)=&\varpi\left[ \ket{00}\bra{00}+\ket{11}\bra{11}\right]+\Sigma\left[\ket{10}\bra{01}+\ket{01}\bra{10}\right]\notag\\&+\Upsilon\left[ \ket{01}\bra{01}+\ket{10}\bra{10}\right],
	\label{rhoout2}
\end{align}
where the elements $\varpi$, $\Sigma$ and $\Upsilon$ are given by
\begin{align*}
	&\varpi=2d\left( a+b\right),\hspace{1cm}\Sigma=2c^{2}e^{-i\varphi}\sin\theta,\\
	&\Upsilon=4d^{2}\cos^{2}\left(\theta/2\right)+\left( a+b\right)\sin^{2}\left(\theta/2\right).
\end{align*}
According to equation (\ref{eq50}), the teleportation fidelity turns out to be
\begin{align}
F\left(\rho_{out}\right)=&\left(\left( \frac{1-\varsigma}{4}\right)^{2}+\frac{\varsigma^{2}\Lambda\left( \xi,\kappa\right) ^{2}}{2}\sin^{2}2\alpha\right) \sin^{2}\theta\notag\\&+\left( \frac{ 1+\varsigma}{4}\right)^{2}\left( 1+\cos^{2}\theta\right),
\end{align}
and subsequently, the average fidelity is provided as
\begin{equation}
F_{av}\left(\rho_{out}\right)=\frac{2}{3}\left( \frac{1+\varsigma}{4}\right)^{2}+\frac{1}{3}\left( \left( \frac{1-\varsigma}{4}\right)^{2}+\frac{\varsigma^{2}\Lambda\left( \xi,\kappa\right) ^{2}}{2}\sin^{2}2\alpha\right).
\end{equation}
\begin{widetext}
	
\begin{figure}[h]
	{{\begin{minipage}[b]{.33\linewidth}
				\centering
				\includegraphics[scale=0.22]{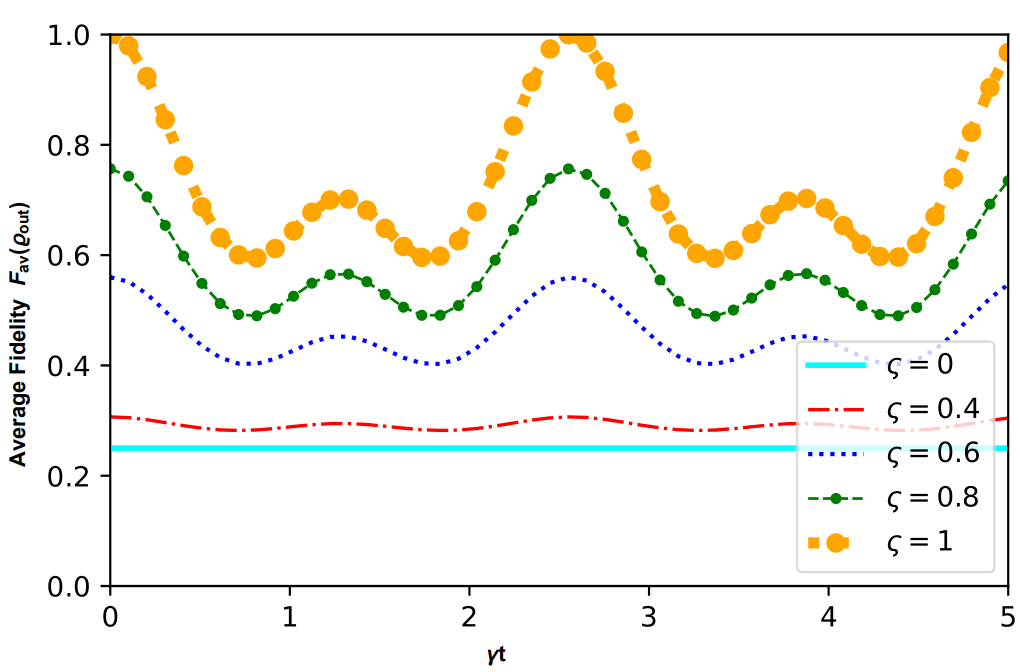} \vfill  $\left(a\right)$ Jaynes-Cummings model
			\end{minipage}\hfill
		\begin{minipage}[b]{.33\linewidth}
			\centering
			\includegraphics[scale=0.22]{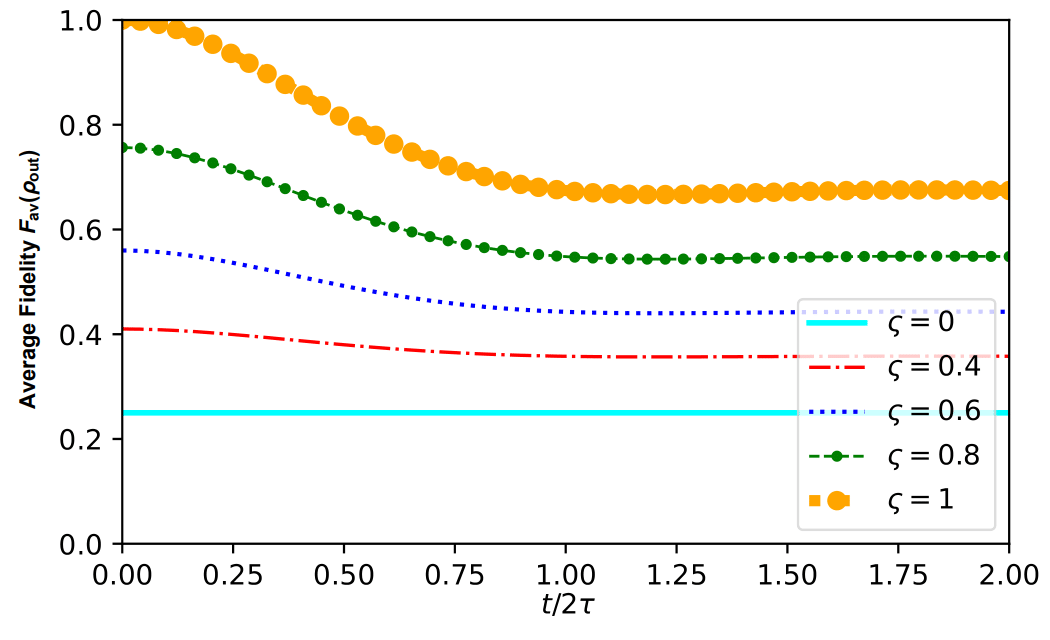} \vfill $\left(b\right)$ Markovian regime
		\end{minipage}\hfill
		\begin{minipage}[b]{.33\linewidth}
			\centering
			\includegraphics[scale=0.22]{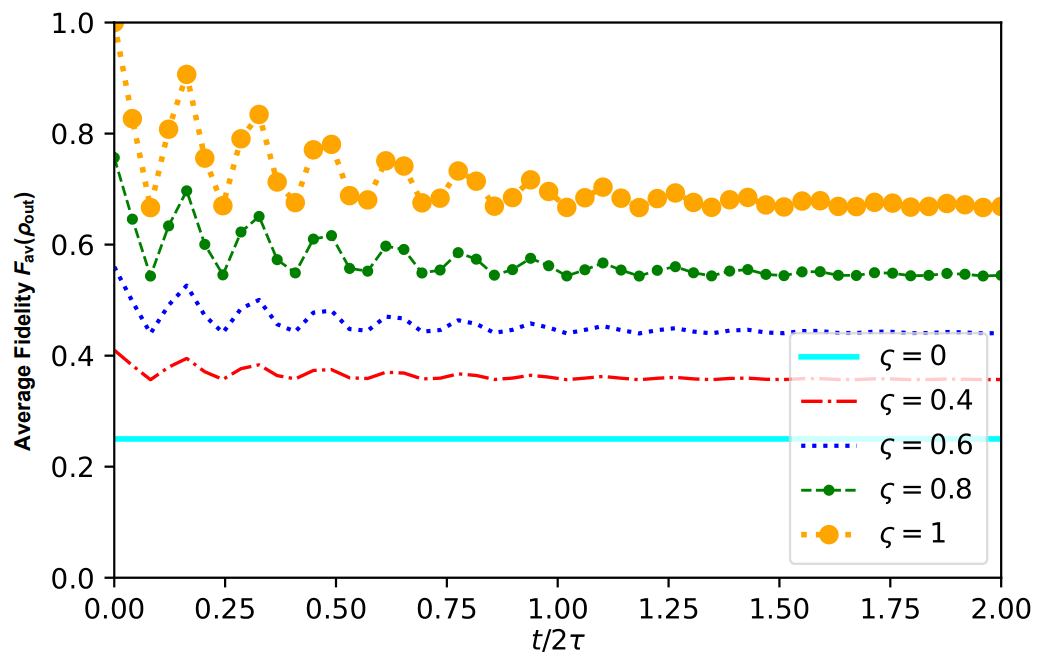} \vfill  $\left(c\right)$ non-Markovian regime
\end{minipage}}}
	\caption{The average fidelity $F_{av}$ as a function of the temps $t/2\tau$ for different values of the purity $\varsigma$ with $\alpha=\pi/4$; panel ($a$) corresponds to the Jaynes-Cumming model, panel ($d$) reflects the Markovian regime for a non-Markovian dephasing model with $a=1$ and $\tau=0.5$, while panel ($c$) for non-Markovian regime with $\tau=5$.}\label{Fig4}
\end{figure}
\end{widetext}
Prior to commencing our analysis of the results displayed in Fig.(\ref{Fig4}), it is crucial to bear in mind some previous investigations. These studies evaluated the effectiveness of quantum teleportation by measuring the average fidelity. A value of $F_{av}<2/3$ indicates a failure in quantum teleportation, $F_{av}>2/3$ signifies an improvement in quantum teleportation, approaching the upper limit achievable through classical protocols. Furthermore, a value of $F_{av}=1$ represents the highest level of teleportation achievement. Fig.(\ref{Fig4}) depicts the behavior of the average fidelity when the resource state that connects Alice and Bob corresponds to the Jaynes-Cumming model represented by state (\ref{rho1}) (i.e., panel ($a$)), as well as when it is a non-Markovian dephasing model described by state (\ref{rho2}) (i.e., panels ($b$)-($c$). In both cases, the teleported state is initially prepared in state (\ref{eq47}), and various values of purity $\varsigma$ are taken into consideration.\par

When Alice and Bob share the Jaynes-Cummings model state, as depicted in Fig.\ref{Fig4}.($a$), we observe a distinct periodic pattern in the average fidelity of the teleported state. This behavior is influenced by the purity of the shared state, denoted by $\varsigma$. As the purity increases, the efficiency of teleportation improves, leading to higher fidelity in the teleported state. Conversely, as the purity decreases, the efficiency of teleportation diminishes, resulting in lower fidelity. To explain further, the average fidelity serves as a measure of how well the quantum information is preserved during teleportation. A value of less than $2/3$ indicates that teleportation fails, and the state is not effectively teleported. On the other hand, when the average fidelity reaches $F_{av}=1$, it signifies a perfect teleportation process, where the teleported state matches the original state with utmost accuracy. By analyzing our results, we can confidently assert that teleportation significantly improves with higher values of purity, as evidenced by the increase in average fidelity. In essence, the higher the purity of the shared state, the better the teleportation outcome. Additionally, referring back to our previous findings, as shown in Fig.(\ref{Fig1}), we notice a similar trend. The average fidelity of the teleported state achieves its maximum when the degree of quantumness (LQFI, LQU, and QC) of the quantum resource state is maximized. In essence, the more quantum properties the resource state possesses, the better the fidelity of the teleported state during the teleportation process. This further supports the significance of quantumness in achieving efficient and accurate quantum teleportation.\par

The results displayed in panels ($b$)-($c$) of Fig.(\ref{Fig4}) depict the evolution of the average fidelity of the teleported state when Alice and Bob share the non-Markovian dephasing model state. We investigate two specific scenarios: the Markovian regime (Fig.\ref{Fig4}$b$) and the non-Markovian regime (Fig.\ref{Fig4}$c$). The study highlights the critical role of state purity in determining the success of quantum teleportation, with pure states demonstrating higher average fidelity compared to mixed states. In the Markovian regime, as seen in Fig.(\ref{Fig4}$b$), the average fidelity shows a gradual decrease in its maximum value over time, eventually stabilizing at a particular level. This behavior suggests that the system exhibits long-term memory. However, the Markovian regime allows for a relatively straightforward transfer of quantum information through teleportation since it does not retain any memory of its previous states. On the other hand, Fig.(\ref{Fig4}$c$) displays the average fidelity of the non-Markovian regime with a specific parameter $\tau=5$, revealing more pronounced short-term fluctuations. The average fidelity exhibits a gradual decrease in its maximum value, followed by an increase after a brief period, reaching a peak lower than the previous one. These fluctuations arise due to transient transitions between different states of the system, leading to temporary variations in average fidelity.

These findings underscore the challenges of transferring quantum information in the non-Markovian regime due to memory effects. The system retains information about its previous states, which can impact the quantumness between particles involved in the teleportation process. Consequently, the memory effects in the non-Markovian regime introduce additional complexity and fluctuations, resulting in a less stable and predictable average fidelity during quantum teleportation. In contrast, the lack of memory in the Markovian regime ensures a smoother and more efficient transfer of quantum information. Accordingly, our analysis shows that state purity, as well as the presence or absence of memory effects, play pivotal roles in influencing the success and efficiency of quantum teleportation. Pure states and Markovian regimes offer more favorable conditions for achieving high average fidelity in the teleported state.

\section{Conclusion}\label{Sec5}
To conclude, we have explored the significance of the quantifiers of non-classical correlations, specifically examining LQU and LQFI, along with quantum coherence. These factors play a crucial role in enhancing our comprehension of quantum information processing and the intricate dynamics within open quantum systems. Our focus was on a comparative analysis of their dynamic characteristics using two distinct physical situations. The first involves a system of two qubits coupled with a single-mode cavity field, while the second centers around the interaction of two qubits with dephasing reservoirs. Our research uncovers that the dynamics of these quantum characteristics are strongly influenced by the initial conditions ($\varsigma$,$\alpha$) and the extent of non-Markovian behavior denoted by $\tau$ within the system. Specifically, our observations highlight that the Jaynes-Cummings model showcases periodic oscillations and distinct sudden-death and revival phenomena, which can be regulated by manipulating the initial state parameters. In contrast, the non-Markovian dephasing model manifests a fundamentally divergent behavior compared to the Markovian system. Within the Markovian context, quantum correlations and coherence prove to be more fragile. Subsequently, we introduced a quantum teleportation scheme founded upon the two models under consideration, in which the resulting two-qubit state from these models is employed as a quantum channel within the framework of a quantum teleportation protocol. The results substantiate that the proposed scheme can be regarded as a reliable protocol for quantum teleportation, utilizing both models as quantum channels.\par




\end{document}